\pdfoutput=1
\documentclass[a4paper]{elsart5p}

\usepackage{graphicx,amssymb,wasysym,natbib}
\usepackage[ansinew]{inputenc}
\usepackage[percent]{overpic}

\begin{document}

\begin{frontmatter}

\title{Exposing metal and silicate charges to electrical discharges:\\Did chondrules form by nebular lightning?}

\author[tubs]{Carsten Güttler\corauthref{cor}},
\ead{c.guettler@tu-bs.de}
\author[tubs]{Torsten Poppe},
\author[ucla]{John T. Wasson},
\author[tubs]{Jürgen Blum}

\address[tubs]{Institut für Geophysik und extraterrestrische Physik,
Technische Universität Braunschweig}
\address[ucla]{Institute of Geophysics and Planetary Physics,
University of California, Los Angeles}

\corauth[cor]{Corresponding author.}

\begin{abstract}
In order to investigate the hypothesis that dust aggregates were
transformed to meteoritic chondrules by nebular lightning, we
exposed silicatic and metallic dust samples to electric discharges
with energies of 120 to 500~J in air at pressures between 10 and
$10^5$ Pa. The target charges consisted of powders of
micrometer-sized particles and had dimensions of mm. The dust
samples generally fragmented leaving the major fraction thermally
unprocessed. A minor part formed sintered aggregates of 50 to
500~$\mu$m. In a few experiments melt spherules having sizes
$\lesssim180\;\mu$m in diameter (and, generally,  interior voids)
were formed; the highest spherule fraction was obtained with
metallic Ni.  Our experiments indicate that chondrule formation by
electric current or by particle bombardment inside a discharge
channel is unlikely.
\end{abstract}


\end{frontmatter}

\section{Introduction}
Chondrules are millimeter-sized spheroids which are common in
primitive chondritic meteorites and formed by melting and
resolidification in the early solar system some 4.56 Myrs ago.
Consisting mainly of FeO- and MgO-rich silicate minerals, most
chondrules are porphyritic, have experienced two or more melting
events, and were incompletely melted in the final melting event
\citep{WassonRubin:2003}. From their abundance and their various
characteristics chondrules provide the key to fascinating insights
into the early solar system.

The process which formed chondrules is still unknown. It must have
been very energetic and is thus in striking contrast to the low
energetic protoplanetary coagulation of dust, presumably taking
place at the same time. Many hypotheses of chondrule formation
processes are listed by \citet{Boss:1996}, and although recent
works favor the formation in nebular shocks or X-wind jets
\citep[see][]{Ciesla:2005}, the formation in nebular lightning
which was originally proposed by \citet{Whipple:1966} is still
under serious discussion.

The idea of the lightning hypothesis is that dust in the solar
nebula underwent tribocharging in non-sticking collisions, and a
subsequent spatial separation lead to large-scale electric fields
which finally caused an electrical breakdown. Inside this
discharge channel preexisting clumps of solid particles were
heated, either by the passage of electric currents, by the
bombardment of surfaces by energetic particles, or by
electromagnetic radiation \citep{HoranyiEtal:1995}. Outside the
channel, radiation emitted by the hot plasma \citep{Wasson:1996,
EisenhourEtal:1994, EisenhourBuseck:1995}, thermal exchange with
the heated gas or recombination of atomic H on the chondrule
surfaces as  compiled by \citet{Desch:2000} may have lead to the
formation of chondrules.

There is extensive former work on the lightning hypothesis
\citep[e.g.][]{MorfillEtal:1993, HoranyiEtal:1995,
GibbardEtal:1997, PilippEtal:1998, DeschCuzzi:2000}, whereas
recent works disregard the idea mainly because cooling times in
this environment are thought to be too fast to form the observed
chondrule textures. Contrariwise, lightning models are supported
by some researchers who have inferred the very cooling times on a
scale of seconds \citep[e.g.][]{GreenwoodHess:1996,
WassonRubin:2003}. Remanent magnetization of chondrules is
furthermore discussed to be a hint that chondrules formed in
regions with high magnetic fields which could result from
lightning \citep{WasilewskiDickinson:2000, ActonEtal:2007}.

Related to our work presented below, it is important that
electrical charge is effectively separated to establish high
potentials for lightnings with sufficient energy to melt
chondrules \citep{DeschCuzzi:2000}. Sufficiently high charge
transfers (not the tribocharge mechanism) were in fact confirmed
by experimental investigations of collision-induced charging of
micrometer sized grains \citep{PoppeSchraepler:2005,
PoppeEtal:2000}.

In this research we followed a procedure similar to that described
by \citet{Wdowiak:1983}, who exposed a mm-sized dust ball of
Allende meteoritic material to a 5 kJ electric discharge. His
experiments left most material unprocessed, produced some sintered
agglomerates, and a few spherules in the size range of 100 to 200
$\mu$m containing bubbles in the interior and on the surface. He
concluded that chondrules did not form by lightning.

In our experiments we exposed dust agglomerates to an electrical
discharge between electrodes. We varied sample material, pressure,
and discharge energy, and we made quantitative analysis of
spherule size distribution and energetic efficiencies of the melt
processes.

Section \ref{setup} describes the experimental setup and all
experimental parameters, and in section \ref{results} the results
are presented. Energetic efficiency, stability of dust aggregates,
the porosity of melt spherules and possible consequences of
aggregate sintering are discussed in section \ref{discussion} and
concluded in section \ref{conclusion}.

\section{Experimental setup}\label{setup}

We used a vacuum chamber suitable for air pressures between 10 and
$10^5$ Pa which contained two electrodes made of silver-coated
copper wires separated by a typical distance of 3~mm. The
electrodes were electrically connected to a 5~$\mu$F capacitor
which could be charged to 7~--~14~kV resulting in a total electric
energy between 123 and 490~J. This setup determined a discharge
duration of approximately 60~$\mu$s as measured with a photodiode.
The maximum light intensity is reached after a few microseconds,
and so is the maximum temperature. This was measured by two-color
photometry which provided an estimation of maximum temperature of
6500~K (emission lines were not taken into account). The ambient
gas in all experiments was air.

\begin{figure}[h]
    \center
    \includegraphics[width=21pc]{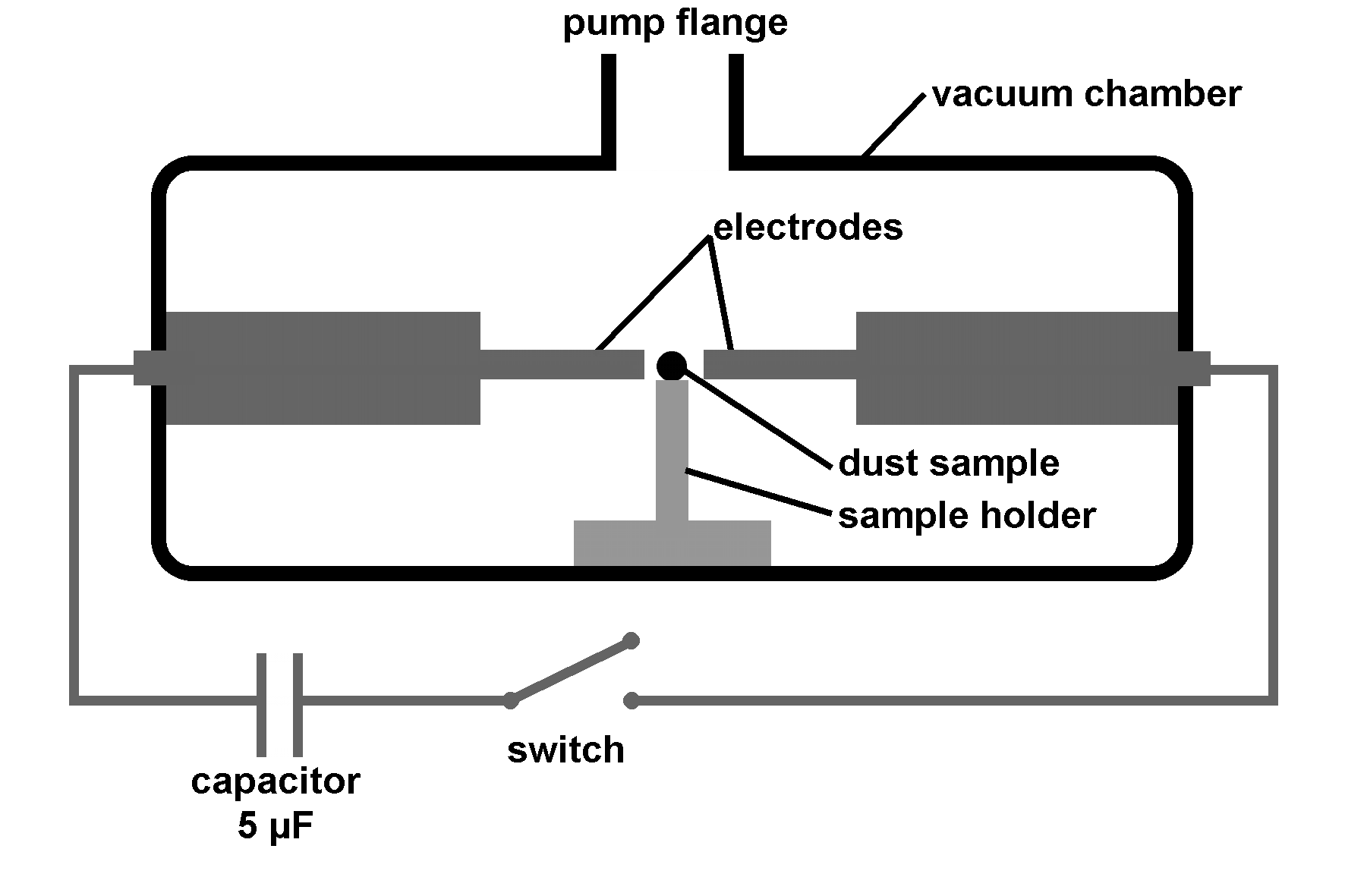}
    \caption{Sketch of the experimental setup. The dust sample is
    placed between two electrodes in a vacuum chamber.}
    \label{sketch}
\end{figure}

Dust powder samples of about 1~mm in size (approx. 1~--~5~mg,
consisting of micrometer-sized grains; cf Table
\ref{material-parameters}) were placed on a plastic (PVC) sample
holder between the electrodes as illustrated in Fig. \ref{sketch}.

To avoid contamination of the sample, the electrodes were replaced
after each experiment. No erosion or melting was observed on the
plastic sample holder and the chamber could easily be cleaned to
avoid contamination with samples of preceding experiments.

\begin{table*}[thb]
    \scriptsize
    \caption{Material parameters of selected dust types. The
    properties for the San Carlos Olivine are calculated by assuming 9
    \% Fayalite and 81 \% Forsterite. Sources: $^a$manufacturer
    information (goodfellow.com), $^b$\citet{HandbookOfPaC},
    $^c$\cite{Wasson:1996}, $^d$ calculated from melting temperature,
    specific heat capacity, and heat of fusion,
    $^e$\citet{PhysPropRocks}}
    \label{material-parameters}
    \begin{tabular*}{\textwidth}{l@{\extracolsep\fill}ccccc}
        \hline
        &nickel&iron&fayalite&San Carlos olivine&silica\\\hline
        density [g cm$^{-3}$]                         &8.9$^a$  &7.87$^a$  &4.30$^b$&3.3$^b$     &2.18 -- 2.65$^b$\\
        melting temperature [°C]                      &1453$^a$ &1535$^a$  &1490$^c$&1861$^{b,c}$&1713$^b$\\
        spec. heat capacity$^b$ [J g$^{-1}$ K$^{-1}$] &0.444    &0.449     &0.652   &0.825       &0.738\\
        heat of fusion [J g$^{-1}$]                   &292$^a$  &272$^a$   &452$^c$ &500$^{b,c}$ &142$^b$\\
        tot. spec. heat$^d$ [J g$^{-1}$]              &928      &952       &1410    &1808        &1391\\
        heat conductivity [W m$^{-1}$ K$^{-1}$]       &90.7$^b$ &80.2$^b$  &3.0$^e$ &5.7$^e$     &1.35$^b$\\
        el. resistivity [$\mu\Omega$$\cdot$cm]        &6.9$^a$  &10.1$^a$  &        &            &\\
        grain size [$\mu$m]                           &0.2 -- 5 &0.1 -- 1.2&$<$ 50  &$<$ 50      &0.9 -- 1.7\\
        mean grain size [$\mu$m]                      &2.33     &0.56      &0.75    &1.43        &1.26\\
        mean axis ratio                               &1.16     &1.15      &1.59    &1.57        &1.15\\\hline
    \end{tabular*}
\end{table*}

For a qualitative analysis, the processed dust sample was
collected on a glass plate on the bottom of the chamber and then
examined with optical or scanning electron microscopy (SEM). Some
target materials formed spherules which could be embedded into
epoxy resin for sectioning. For quantitative analysis of the melt
spherules, the sample was collected on a sheet of paper and then
put into a mold of $1\times1$~cm. This mold was then scanned by
light microscopy for counting the spherules and measuring their
sizes.

Charges were composed of fayalite, forsteritic olivine, SiO$_2$,
iron (oxidized on the surface), and nickel as described in Table
\ref{material-parameters}. Some studies were carried out on
peridotite 699 collected by G. Witt-Eickschen at Sch{\"o}nfeld,
West Eifel, Germany.  It has the following chemical composition:
SiO$_2$, 44.14~\%; Al$_2$O$_3$, 3.07~\%; FeO, 7.17~\%; MgO,
40.93~\%; CaO, 2.8~\% and Na$_2$O, 0.29~\%. The Mg/Mg+Fe molar
ratio is 0.912. Other similar samples were discussed in the paper
by \citet{WittSeck:1987}. The choice of materials was lead by the
idea of clean and defined conditions to enable repeatability and
moreover the peridotite is directly compositionally relevant.
Furthermore, the conductivity of nickel and iron must play a role
in electric discharge and the SiO$_2$ monomers have already been
used in sintering experiments \citep{Poppe:2003} and are well
analyzed \citep{HeimEtal:1999} to permit calculations.

\section{Results}\label{results}

\begin{figure*}[t]
    \center
    \includegraphics[width=\textwidth]{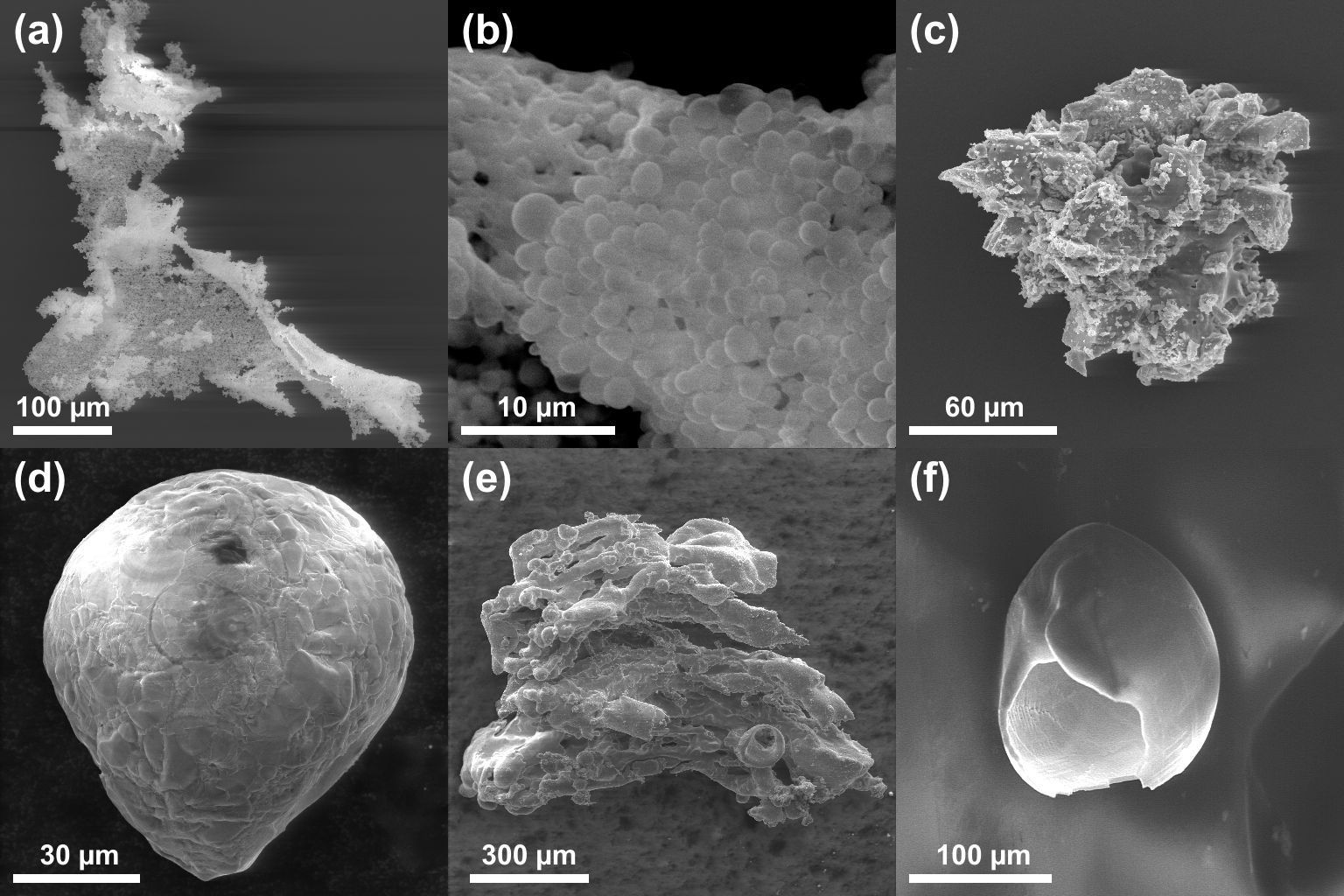}
    \caption{Typical experimental outcomes: (a) Sintered agglomerate
    of SiO$_2$ monomers, (b) magnified SiO$_2$ agglomerate, (c)
    sintered agglomerate of fayalite grains, (d) nickel spherule (e)
    iron agglomerate and (f) hollow iron sphere.}
    \label{result-image}
\end{figure*}

\subsection{Phenomenology of processed material}
Instantly after triggering the discharge, the dust sample
exploded. This happened at all pressures, discharge energies, and
sample materials. However, small parts of the samples were
thermally processed and could afterwards be found on the glass
plate. Apart from unprocessed, dispersed dust, sintered dust
agglomerates (50~--~500~$\mu$m in diameter) were common and,
depending on the material, also spherules (diameter
$\lesssim180\;\mu$m) were found. Iron powder showed the most
product variety including hollow spheres and millimeter-sized,
comparatively dense agglomerates. Figure \ref{result-image} shows
a variety of processed dust samples found after discharge heating.
Experimental parameters and outcomes are compiled in Table
\ref{result-table}. To guarantee reproducibility, all specified
experiments were conducted at least three times. Hence, the
results presented in the following are representative for a series
of experiments with same materials and parameters.

\begin{table*}
    \scriptsize
    \caption{Experiment results for different parameters. If spherules
    are formed, agglomerates of sintered (``sintered'') grains are
    also common. Exception: Iron charges formed dense aggregates which
    do not resemble the original grains (''big agglomerate'').
    Unprocessed dust was always present.}
    \label{result-table}
    \begin{tabular*}{\textwidth}{l@{\extracolsep\fill}cccccc}
        \hline
        discharge energy, gas pressure&iron&peridotite 699&fayalite&olivine&silica&nickel\\\hline
        &big agglomerate,&&porous&&&porous\\
        \raisebox{1.5ex}[-1.5ex]{456 J, $10^5$ Pa}&hollow spherules&&spherules&\raisebox{1.5ex}[-1.5ex]{sintered}&\raisebox{1.5ex}[-1.5ex]{sintered}&spherules\\
        456 J, $10^3$ Pa&spherules&sintered&&&sintered&\\
        141 J, 10 Pa&spherules&sintered&sintered&sintered&&sintered\\
        456 J, 10 Pa&spherules&sintered&sintered&sintered&sintered&sintered\\
        \hline
    \end{tabular*}
\end{table*}

Sintered agglomerates like Fig. \ref{result-image} (a) -- (c)
formed from silicates and nickel with all experimental parameters.
Figure \ref{result-image} (b) indicates that the energy
transferred to the sample was insufficient to completely melt the
SiO$_2$ powder. The original 1 $\mu$m silica spheres were sintered
to a varying extend as can be concluded from similar structures
found in sintering experiments by \citet{Poppe:2003}. Due to
preparation in a ball mill, the fayalite grains varied much more
in size and shape but sintering is nevertheless suggested from
Fig. \ref{result-image} (c) which shows an aggregate partly
consisting of rounded shapes and partly of grains with original
sharp edges and of original grain size. For all parameters tested
(Table \ref{result-table}), peridotite and olivine powder only
formed sintered agglomerates looking similar to the fayalite
agglomerate in Fig. \ref{result-image} (c). Some spherules were
found in experiments with nickel and fayalite powder at $10^5$~Pa
and also for iron at pressures below $10^3$~Pa. Due to their
similarity only one exemplary nickel spherule is presented in Fig.
\ref{result-image} (d). For pressures below $10^5$~Pa, nickel
powder only formed sintered aggregates $\lesssim80\;\mu$m in size,
resembling the SiO$_2$ aggregate in Fig. \ref{result-image} (a)
and (b). In each experiment with iron powder at $10^5$~Pa gas
pressure, one or two dense agglomerates of $\sim$1~mm were formed
as shown in Fig. \ref{result-image} (e). These agglomerates did
not form by sintering the original iron grains as can be seen from
the thickness of the substructures ($\sim$30~$\mu$m) which are
larger than the original grain size (spheres of
0.1~--~1.2~$\mu$m). Figure \ref{result-image} (f) shows a hollow
sphere of $\sim150\;\mu$m with a wall thickness of only some
10~$\mu$m, which formed from iron powder together with those dense
agglomerates in Fig. \ref{result-image} (e).

\subsection{Quantitative analysis of spherules}
\begin{figure}[h]
    \center
    \includegraphics[width=70mm]{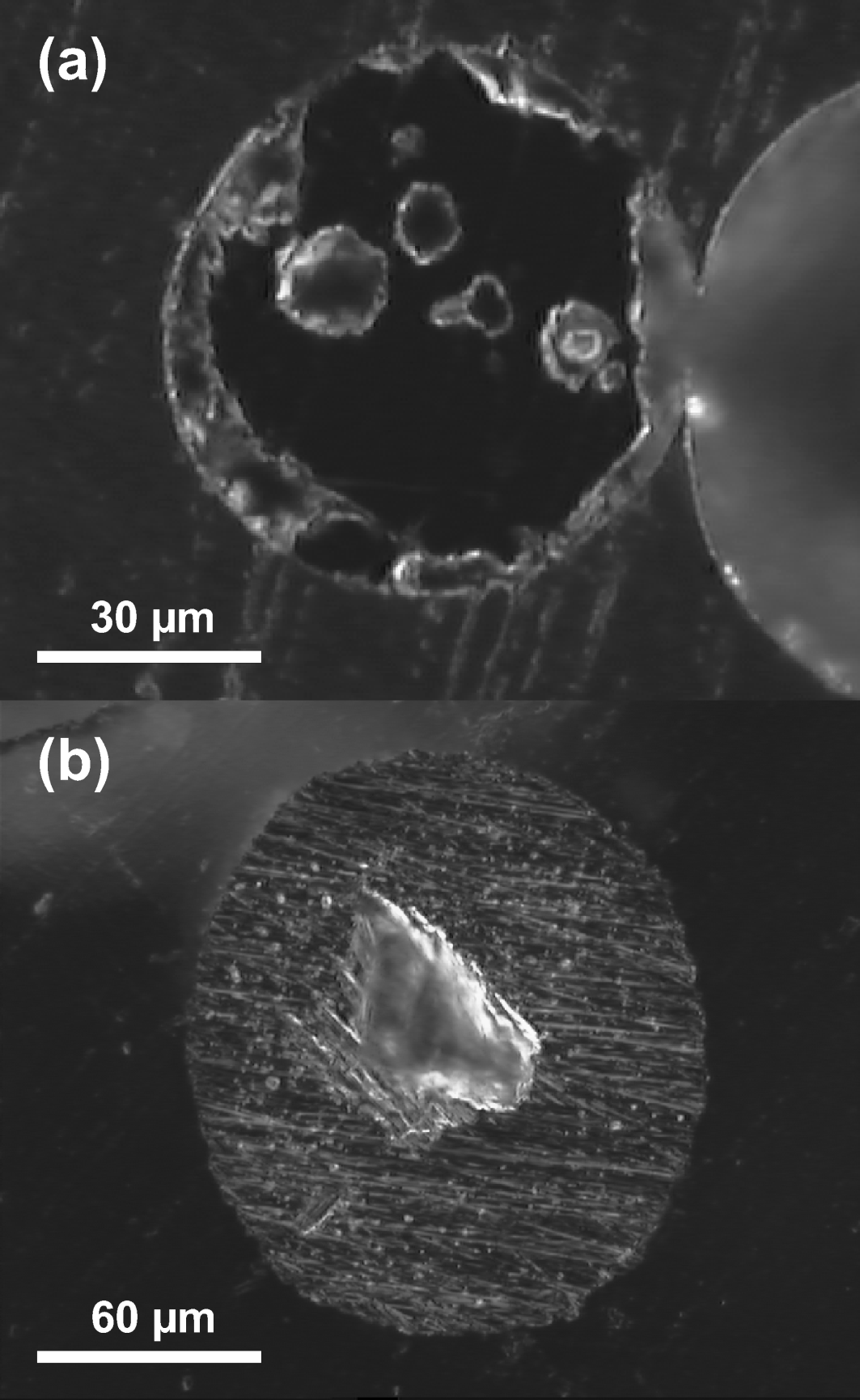}
    \caption{Embedded and sectioned spherules made at $10^5$~Pa and
    490~J. (a) Fayalite spherule with a measured porosity of 14~\% and
    (b) nickel spherule with a porosity of 11~\%.}
    \label{sections}
\end{figure}

For some of our experimental parameters, spherules were found.
Nickel and fayalite formed spherules at $10^5$~Pa air pressure as
shown in Fig. \ref{result-image} (d) for nickel. Fayalite as well
as nickel spherules show a varying porosity from 0 to 60~\% (cf.
Fig. \ref{sections}). The mean porosity computed from 11 fayalite
spherules and 19 nickel spherules was $49\pm4$~\% and
$5.2\pm1.7$~\%, respectively. Solid iron spherules were found for
pressures $\lesssim10^3$~Pa whereas all iron spherules formed at
$10^5$~Pa were hollow bubbles as can be seen from the example in
the SEM image in Fig. \ref{result-image} (f).

\begin{figure}[t]
    \center
    \includegraphics[height=21pc,angle=90]{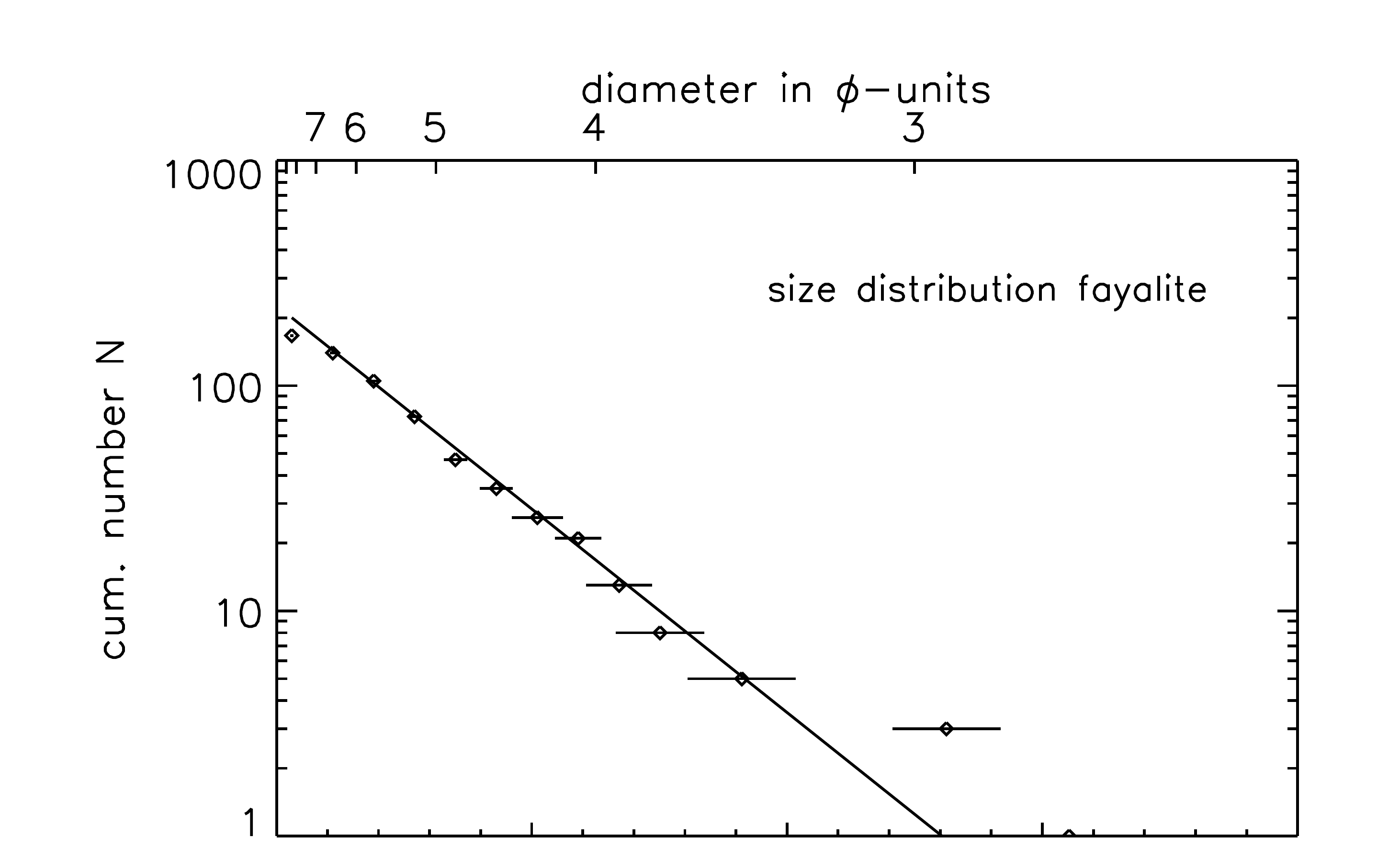}
    \includegraphics[height=21pc,angle=90]{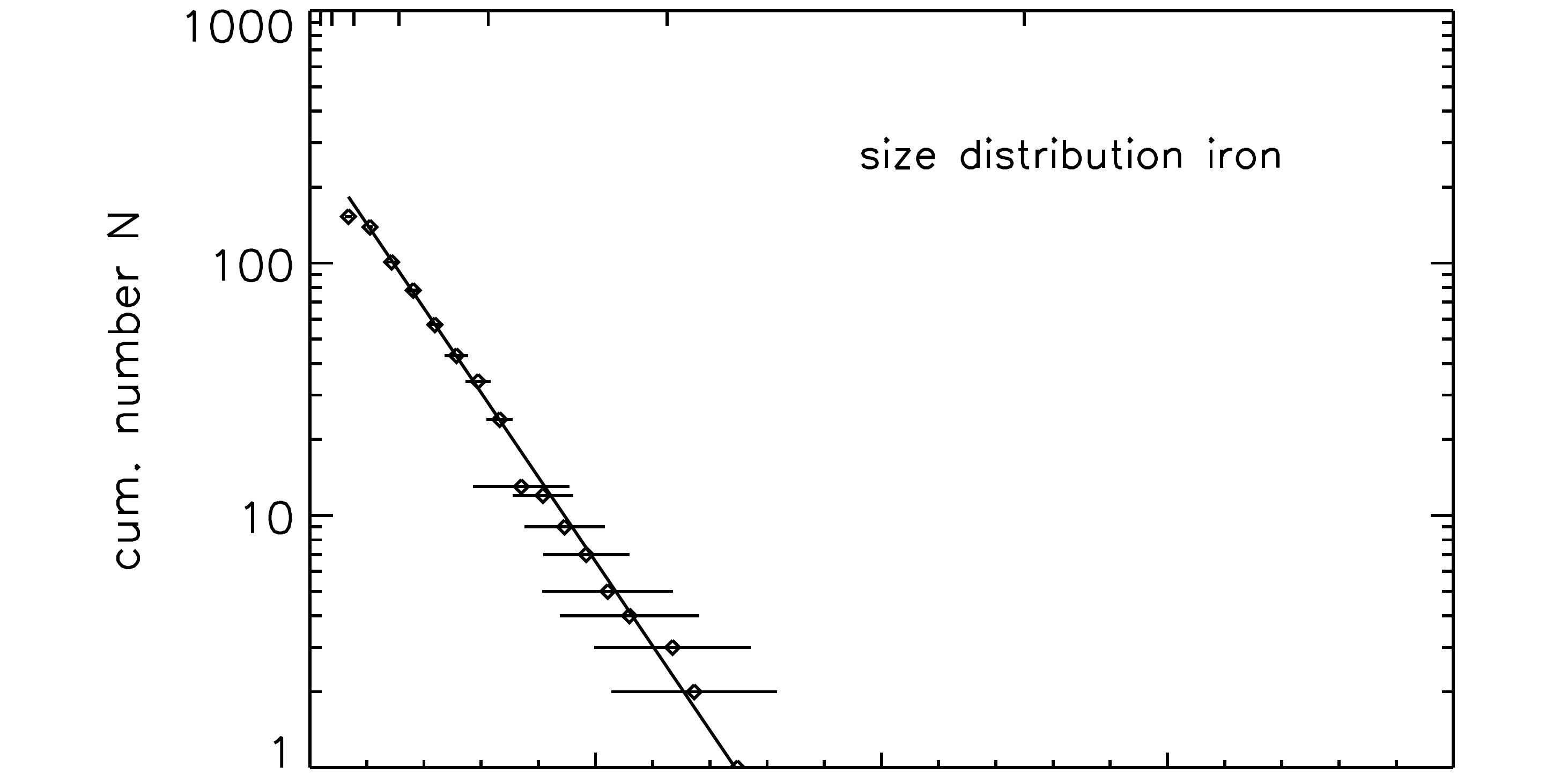}
    \includegraphics[height=21pc,angle=90]{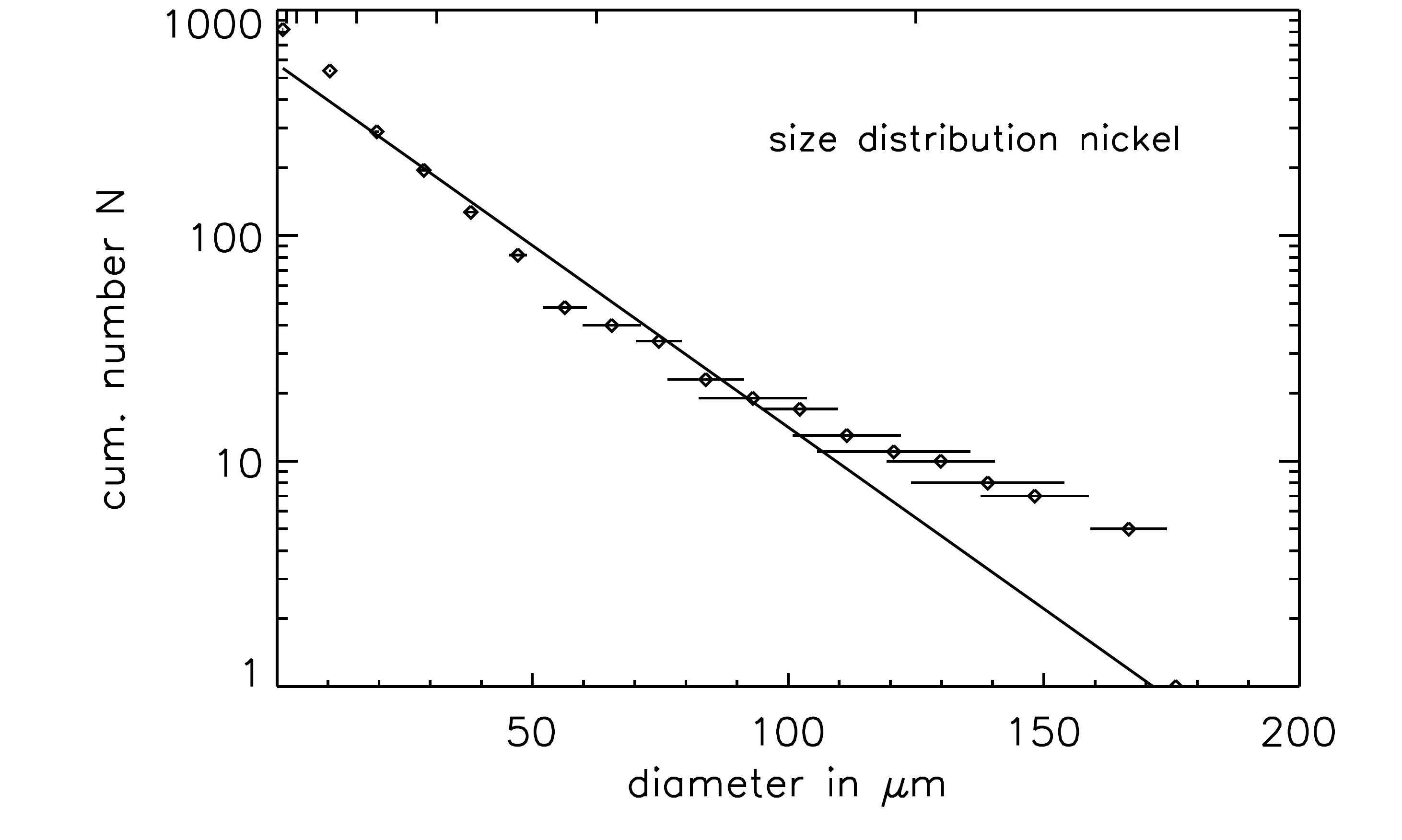}
    \caption{Cumulative size distribution for all counted spherules in
    one experiment. From top to bottom: Fayalite, iron, and nickel.}
    \label{size-distribution}
\end{figure}

In addition to examining single spherules, the number and size
distribution of spherules produced in individual experiments were
determined for the experiments in which spherules formed. The
sizes were binned, cumulated, and plotted with a logarithmic
number-axis (Fig. \ref{size-distribution}). The resulting straight
line suggests an exponential behavior $\tilde{N}(R)=a\cdot
\exp(-bR)$, where $R$ is the spherule radius, $\tilde{N}$ is the
number of counted spherules of radius $\geq$ $R(\tilde{N})$ and
$a$ and $b$ are fitting parameters. Figure 4 shows these size
distributions for all three conditions under which spherules were
found. The range of spherule sizes used for the fit was limited to
spherules $>5\;\mu$m because smaller spherules were poorly
resolved on the images. The value $R_{max}=\frac{3}{b}$ defines
the radius of spherules which contain the most mass for one given
distribution because it gives the maximum of the function
$N(R)\sim R^3\cdot e^{-bR}$. This radius was found to be
approximately 40~$\mu$m for fayalite and nickel spherules and 16
$\mu$m for iron spherules (cf. Table \ref{mass-size-table}).

\begin{table}
    \scriptsize
    \caption{Total mass in spherules and $R_{max}$ for all experiments
    in which the spherules were counted (discharge energy 456
    J).}\vspace{1mm}
    \label{mass-size-table}
    \begin{tabular*}{21pc}{l@{\extracolsep\fill}ccc}
        \hline
        material & pressure & total mass & $R_{max}$\\
        & [Pa] & [$\mu$g] & [$\mu$m]\\
        \hline
        fayalite & $10^5$ &  23.6 & 44.2\\
        fayalite & $10^5$ &  38.9 & 36.0\\
        iron     & 10     &  18.2 & 20.9\\
        iron     & 10     &   4.6 & 13.1\\
        iron     & 10     &  13.4 & 19.4\\
        iron     & 10     &   9.9 & 12.4\\
        nickel  &  $10^5$ & 319.6 & 40.4\\
        \hline
    \end{tabular*}
\end{table}

Furthermore the total mass transformed into spherules was
calculated by summing the mass of all counted spherules, taking
into account the measured mean porosity depending on the material.
According to this, the mass that was transformed into spherules is
31~$\mu$g for fayalite, 12~$\mu$g for iron, and 320 $\mu$g for
nickel powder (cf. Table \ref{mass-size-table}). The total energy
for melting these masses can be deduced using Table
\ref{material-parameters}. It is thus possible to provide an
energetic efficiency of discharge melting which is the fraction of
energy consumed for melting divided by the input electrical
discharge energy. We find an energetic efficiency of 0.006~\%
(fayalite), 0.004~\% (iron) and 0.06~\% (nickel).

\subsection{Pre-sintered aggregates in the discharge}
In order to assess the force driving the fragmentation,
pre-sintered dust aggregates of SiO$_2$ powder have been put into
the discharge. These were similar to aggregates described by
\citet{Poppe:2003} and were sintered in a furnace for one hour at
1050°C and 1100°C, respectively.

The agglomerates which were sintered at 1050°C fragmented, were
completely dispersed, and were thermally unprocessed whereas those
sintered at 1100°C were stable but did also not melt. The tensile
strength of a macroscopic sintered aggregate is unknown but the
neck radius between single monomers, giving an indication for the
bondings, was measured by \citet{Poppe:2003} to 0.4 (1050°C) and
0.55 (1100°C) of the particle radius.

\section{Discussion}\label{discussion}
\subsection{Agglomerate destruction and melt spherule sizes}
Under all experimental conditions, most of the dust sample was
dispersed rather than thermally processed. \citet{Wdowiak:1983}
made similar observations but, unfortunately, did not provide
details. We exclude electrical forces as cause of dust
fragmentation (see appendix \ref{destruction-calculation}).
Instead, we attribute the explosion to the expansion of the
discharge channel, possibly supported by evaporation of volatile
materials, both processes dependent on the discharge temperature.
There is no indication that hypothetical nebular lightning
discharges had lower temperatures than our experimental discharge,
in most models much higher temperatures are inferred
\citep{HoranyiEtal:1995}. Thus, nebular lightning should have been
even more violent than our experiments. So in nebular lightning,
loosely-bound aggregates of micrometer-sized grains would not have
survived within the discharge channel. If spherules were formed,
the heating must thus have been by above mentioned processes
outside the high-temperature channel.

Nevertheless, a few melt spherules were produced in our
experiments, but they amount a very small fraction of the original
material and are, in most cases, smaller than chondrules. It is
thus obvious that some small fragments of the original sample were
melted. Because we know so little about nebular lightning and the
various processes by which it could cause dust to melt, it is
difficult to generalize further.

\subsection{Energetic efficiency}
While the question whether nebular lightning could have had as
much energy as necessary to melt chondrules has been treated
\citep{DeschCuzzi:2000}, the question how effective lightning
energy is for melting has not yet been discussed. The ultimate
energy source for a possible chondrule melting by lightning is
potential energy of disk material in the gravitational field of
the sun. At 2~AU, this specific energy is $E=\gamma
M_{\astrosun}/r=4\cdot10^8$~J~kg$^{-1}$ whereas the required
energy for the complete fusion of silicates like fayalite is at
least $1.5\cdot10^6$~J~kg$^{-1}$ (cf. Table
\ref{material-parameters}). Taking into account at least two
heating cycles \citep{Wasson:1993, Wasson:1996} this leads to a
lower limit of an efficiency of $\sim1$~\% necessary to melt
chondrules under the unrealistically favorable condition that all
potential energy is converted into electric lightning energy. Even
then, the efficiency found in the experiments is lower by more
than one order of magnitude, and this is only true for nickel at
$10^5$~Pa (fayalite and iron two orders of magnitude). So the
energetic efficiency is much too low in order to explain chondrule
formation by particle bombardment heating or electrical current
heating inside nebular lightning channels. However, our results do
not exclude other heating mechanisms as discussed in introduction.
E.g., nebular lightning heating by electromagnetic radiation could
have been stronger than in our experiments due to different length
scales. A possible step forward to treat the problem of radiation
heating could be experiments on dust aggregate melting by laser
radiation (Springborn et al. in preparation).

We find a clear trend that conductive materials are more easily
melted in the experimental discharge than nonconductive: Nickel
has the highest energetic efficiency of all used powders and apart
from the moderate efficiency in forming solid spherules, also the
iron powder shows a high processing rate in forming one big
agglomerate. Fayalite is the only nonconducting powder which
melted at all. Possible reasons for the greater melting of the
conductors may be the heating by interior currents and the
focusing of electric flux lines followed by an increased number of
bombarding charged particles. From this we conclude that the
conductive fraction of chondrule precursor material would have
played an important role in a formation process with heating by
nebular lightning.

\subsection{Porosity}
Most nickel and fayalite spherules were porous. In fayalite
spherules bubbles made up 49~\% of volume on average, and, in
nickel spherules it was 5~\%. Since bubbles are rarely observed in
chondrules their presence in our products seems to be an argument
against chondrule formation in nebular discharges, as also noted
by \citet{Wdowiak:1983}. However, this argument is not strong as
it has been recognized that bubble retention could be a
consequence of low energetic efficiency which resulted in a low
degree of melting. \citet{MaharajEtal:1993} argue that bubbles are
typical for a low degree of melting because viscosity is
temperature dependent such that at low temperatures bubbles cannot
escape as easily as at high temperatures. This is consistent with
the finding that fayalite has a lower energetic efficiency than
nickel and also a higher porosity. A higher energetic efficiency
might therefore lead to a higher degree of melting and hence to a
lower porosity.

\subsection{Other aspects}
We found that experiments more often resulted in sintering and
strengthening of agglomerate fragments than in melt spherule
formation. Therefore, other effects of possible nebular lightning
on pristine solar nebula material should also be considered. An
obvious link could exist to protoplanetary dust aggregation for
which the collisional behavior, and thus the strength, of growing
agglomerates is a key question \citep{BlumWurm:2008}. Collision
experiments with well defined sintered dust agglomerates
\citep{Poppe:2003} are ongoing to study their behavior (Krause and
Blum, pers. comm.). In turn, also the search for features of
sintering in primitive solar system material could give
indications wether lightning occurred in the solar nebula or not.

\section{Conclusion}\label{conclusion}
We performed experiments on electrical discharge heating of
micrometer-sized silicatic and metallic dust powder in order to
investigate the hypothesis that chondrules may have formed by
lightning in the solar nebula. We found that discharge heating is
energetically extremely inefficient in our experiments. Dust
aggregates were destroyed instead of transformed into melt
spherules. The few spherules produced were smaller than meteoritic
chondrules and, in contrast to them, much more porous.

While it is true that some spherules had the size of small
chondrules and porosity may be a consequence of low energetic
efficiency, aggregate destruction and low energetic efficiency are
strong and independent arguments against chondrule formation
inside nebular lightning channels. Any theory on chondrule
formation in nebular lightning discharges must explain why nebular
lightning did not destroy chondrule precursor aggregates and must
also deal with the low energetic efficiency. However, other
lightning-related heating mechanisms outside the lightning channel
are not excluded by the results of our experiments.

The experiments showed however that if lightning occurs in the
early solar system at all, it will be able to alter dust
agglomerates with possible consequences to protoplanetary dust
growth.

\begin{appendix}
\section{Estimation of agglomerate destruction effects}\label{destruction-calculation}
The order of magnitude for thinkable electric effects which can
lead to the fragmentation of a dust aggregate are hereafter
estimated: \emph{(i)} Electrostatic repulsion and \emph{(ii)}
forces due to field gradients will be calculated for 1 $\mu$m
SiO$_2$ monomers because of their well-known material parameters
\citep{HeimEtal:1999,BlumSchraepler:2004}.

\emph{(i)} The Coulomb force between two charged grains in contact
is $F=(Ze)^2/(4\pi\varepsilon_0D^2)$, where $Z$ is the number of
electrons of charge $e$ on each grain and $D=1\;\mu$m is the
distance of the grain centers. The charge on each grain must be
17,000 elementary charges to overcome the adhesive force of 67 nN
measured by \citet{HeimEtal:1999} and lessens only slightly for a
chain of charged grains.

\emph{(ii)} If a chain of SiO$_2$ spheres is equally charged, the
force due to the difference in the electric field on the one end
to the other may break it. This happens when the repulsive force
$F=ZeE'D\frac{n(n+1)}{2}$ is stronger than the adhesive force of
67 nN, where $n$ is the number of monomers in a chain and $E'$ is
the gradient in the electric field in V/m$^2$. One case in the
experiment with the lowest field gradient in which the sample
still fragmented is for an electrode distance of 35 mm and a
potential difference of 13.5 kV between the electrodes, which
results in a field gradient of $10^5$ V/m$^2$ at the edge of the
dust sample. Assuming a chain of 10 monomers on the surface which
are equally charged, $8\cdot10^{10}$ elementary charges had to be
deposited on each monomer to overcome the adhesive force of 67 nN.

\citet{HoranyiEtal:1995} calculated the charge on a mm-sized dust
ball in a 30 eV hot plasma to $5\cdot10^7$ electrons, which leads
to only 10 charges per monomer assuming that the aggregate is
porous and only surface layers are charged. This is much too
little for both electrical effects.
\end{appendix}

\section*{Acknowledgement}
We thank Tilman Springborn for his support and Dominik Hezel for
valuable discussion and acquisition of the peridotite. This work
was funded by DFG (Deutsche Forschungsgemeinschaft) under grant PO
817/2.

\bibliography{literature}

\end{document}